# Conversion of out-of-phase to in-phase order in coupled laser arrays with second harmonics

Chene Tradonsky,* Micha Nixon, Eitan Ronen, Vishwa Pal, Ronen Chriki,
Asher A. Friesem and Nir Davidson

*Department of Physics of Complex Systems, Weizmann Institute of Science, Rehovot 7610001, Israel*
*Corresponding author: chene.tradonsky@weizmann.ac.il*

A novel method for converting an array of out-of-phase lasers into one of in-phase lasers that can be tightly focused is presented. The method exploits second harmonic generation and can be adapted for different laser arrays geometries. Experimental and calculated results, presented for negatively coupled lasers formed in a square, honeycomb, and triangular geometries are in good agreement.

## 1. INTRODUCTION

Phase locked laser arrays, where all the lasers have a common frequency and a constant phase difference relative to each other, have interesting physical properties that can be exploited for several applications. For example, phase locked laser arrays can serve as excellent tools to investigate the properties of coupled oscillators as analogue to coupled classical-spins arrays [1], the behavior of frustrated array configurations [2], the conditions for passive phase locking and phase locking dynamics [1,3,4], and the properties of complex networks [5–7]. Also, phase locking of many lasers can result in a combined laser whose output can be highly focused [8–10].

Phase locking can be achieved with different coupling techniques [11,12]. These include evanescent coupling [13,14], antiguided coupling [15,16], mirrors coupling [17], diffractive coupling with Talbot (or fractional Talbot) cavity [18–20], diffractive coupling with self-Fourier cavity [21,22], and diffractive coupling with spatial filtering inside degenerate cavity [23–25]. In general, these coupling techniques can result in either positive coupling that tends to lock neighboring lasers with a zero phase shift between them (in-phase locking) [12,14] or negative coupling that tends to lock neighboring lasers with a $\pi$ phase shift between them (out-of-phase locking) [12–14]. The arrangements for obtaining out-of-phase locked lasers are typically simpler, with fewer elements and less loss (especially in small arrays) than those for obtaining in-phase locked lasers [11,20]. Also, there are some laser array geometries (for example honeycomb geometry) in which it is very difficult to lock the lasers in-phase [11]. Unfortunately, out-of-phase locked arrays have a super mode output that cannot be tightly focused. Thus, it is necessary to convert these out-of-phase lasers to a super mode of in-phase lasers in order to obtain tight focusing, as was already achieved by means of specialized elements such as phase shifters [11,26,27].

Here we present a novel and relatively simple method for converting an array of out-of-phase lasers to one of in-phase lasers by resorting to second harmonic generation (SHG). The SHG results in phase doubling [28–30], so all participating laser outputs, whether in-phase or out-of-phase, will be in-phase. The use of SHG is particularly attractive when newer wavelengths (e.g. visible rather than near infra-red) are needed. Specifically, incorporating second harmonic generation with a phase locked array of lasers could lead to a combined high power with good beam quality in the visible region. Moreover, as we will show, the second harmonics reveal hitherto unknown properties of some laser array geometries (for example triangular geometry) which cannot be detected by the first harmonics intensity distribution. In this method we exploit a degenerate cavity laser to form an array of coupled lasers and a commercially available nonlinear crystal to obtain SHG. We performed experiments and calculations on two coupled lasers and then on arrays of coupled lasers formed in square, honeycomb and triangular geometries.

## 2. EXPERIMENTAL ARRANGEMENT

The arrangement for obtaining an array of negatively coupled phase locked lasers, forming first and second harmonics and their detection is shown schematically in Fig. 1. It consists of three main parts. The first part contains a degenerate cavity laser configuration [2,31], where many independent lasers are formed and coupled [32]. It is comprised of two lenses in a 4f telescope, a 10 mm wide Nd-YAG gain medium, a high reflecting rear mirror at one end and a mask of apertures and a 95% reflectivity output coupler at the other end. The 4f telescope assures that any field distribution will be reimaged after a round trip in the cavity. Accordingly, the mask forms an array of individual lasers with specific lattice geometry. Diffractive coupling between adjacent lasers is introduced by shifting the output coupler away from the mask. The distance between the mask and the output coupler determines the strength and sign of coupling. For negative coupling, that distance is a quarter of the Talbot length (¼zT) [20,33]. The second part includes a second telescope which focuses the output light from the degenerate cavity laser onto a 5 mm cube KTP type two nonlinear crystal to obtain the second harmonic. The third part contains an arrangement of lenses and filters for detecting the near-fields and far-fields of the first and second harmonics.

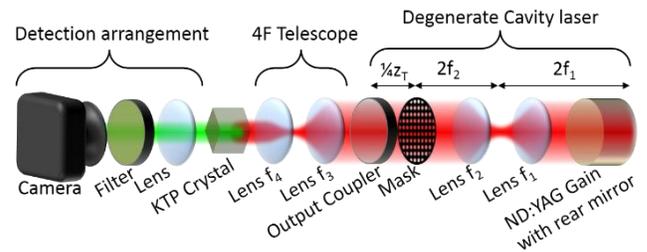

Fig. 1. The arrangement for obtaining an array of negatively coupled phase locked lasers, forming first and second harmonics and their detection. It consists of three main parts: a degenerate cavity laser with an intra-cavity mask of holes so as to obtain a desired laser array geometry, a telescope with a KTP nonlinear crystal to obtain second harmonic, and an arrangement to detect the near-field and far-field of the first and second harmonics.

## 3. RESULTS AND DISCUSSION

In order to demonstrate the principle of converting out-of-phase lasers ($\pi$ phase shift between neighboring lasers) into in-phase lasers ($2\pi$ phase shift) by means of SHG, we began with two coupled lasers, using the arrangement shown in Fig. 1 and a mask of two holes. The illustration for phase shift doubling and experimental intensity distributions for the near and far fields of first and second harmonics are presented in Fig. 2. As expected, the distance between the lobes in the far field intensity distribution of the second harmonic is half that of the first harmonic. Moreover, the darkness in the center of the first harmonic indicates out-of-phase locking ($\pi$ phase shift), and the bright lobe in the center of the second harmonic indicates in-phase locking ($2\pi$ phase shift).

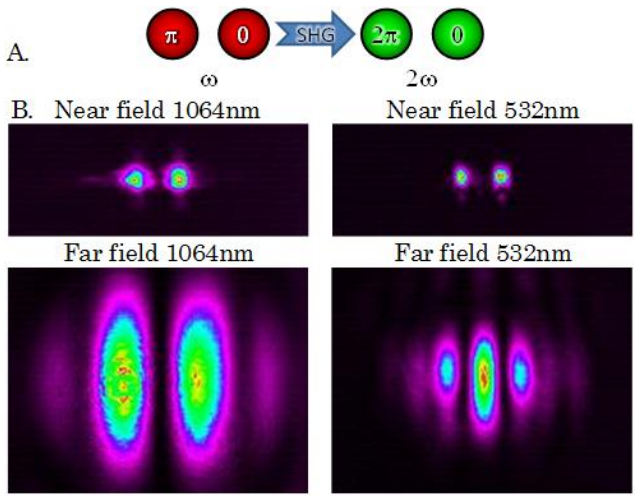

Fig. 2. Two coupled lasers. A. Schematic illustration of phase shift doubling in the second harmonics, where an out-of-phase locking ($\pi$ phase shift) in the first harmonic is converted into an in-phase locking ($2\pi$ phase shift) in the second harmonic. B. Experimental intensity distributions for the near and far fields of first and second harmonics.

We then proceeded to an array of 334 coupled lasers formed in square geometry, using a mask with 334 holes. The illustration for phase shift doubling and experimental intensity distributions for the near and far fields of first and second harmonics are presented in Fig. 3. Here again the distance between the lobes in the far field intensity distribution of the second harmonic is half that of the first harmonic. The darkness in the center of the first harmonic indicates out-of-phase locking ($\pi$ phase shift between neighboring lasers), and the bright lobe in the center of the second harmonic indicates in-phase locking ($2\pi$ phase shift). Finally, the sharp peaks in the far field intensity distributions indicate long range order without tight focusing for the first harmonic, but with tight focusing for the second harmonic.

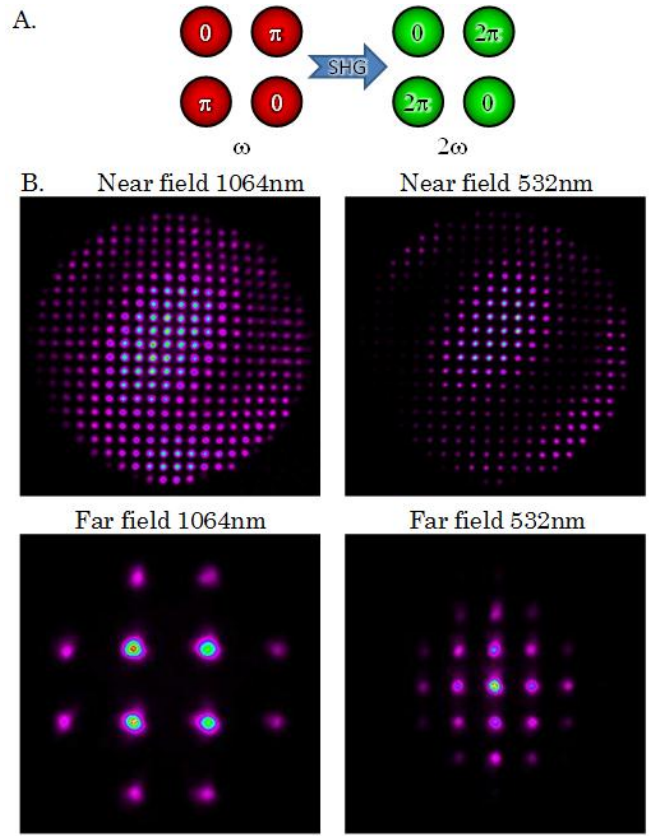

Fig. 3. An array of 334 coupled lasers formed in a square geometry. A. Schematic illustration of phase shift doubling in the second harmonics, where an out-of-phase locking ($\pi$ phase shift) in the first harmonic is converted into an in-phase locking ($2\pi$ phase shift) in the second harmonic. B. Experimental intensity distributions for the near and far fields of first and second harmonics.

We also investigated an array of 199 coupled lasers formed in honeycomb geometry, using a mask of 199 holes. The illustration for phase shift doubling and experimental and numerically calculated intensity distributions for the near and far fields of first and second harmonics are presented in Fig. 4. Here again the distance between the lobes in the far field intensity distribution of the second harmonic is half that of the first harmonic. The darkness in the center in the first harmonic indicates out-of-phase locking ($\pi$ phase shift between neighboring lasers), and the bright lobe in the center of the second harmonic indicates in-phase locking ($2\pi$ phase shift). Finally, the sharp peaks in the far field intensity distributions indicate long range order without tight focusing for the first harmonic, but with tight focusing for the second harmonic.

Finally, we investigated an array of 299 negatively coupled lasers formed in triangular geometry. In such geometry it is impossible to obtain a phase difference of $\pi$ between neighboring lasers. Indeed, when the lasers are phase locked, the phases of neighboring lasers are either 0, $+2\pi/3$, $-2\pi/3$ or 0, $-2\pi/3$, $+2\pi/3$ [11,12]. Specifically, a bistable array is formed with either vortex or antivortex phase distribution. In such an array a simple doubling of the phases will convert a vortex phase distribution into an antivortex phase distribution and vice versa, i.e. the far-field

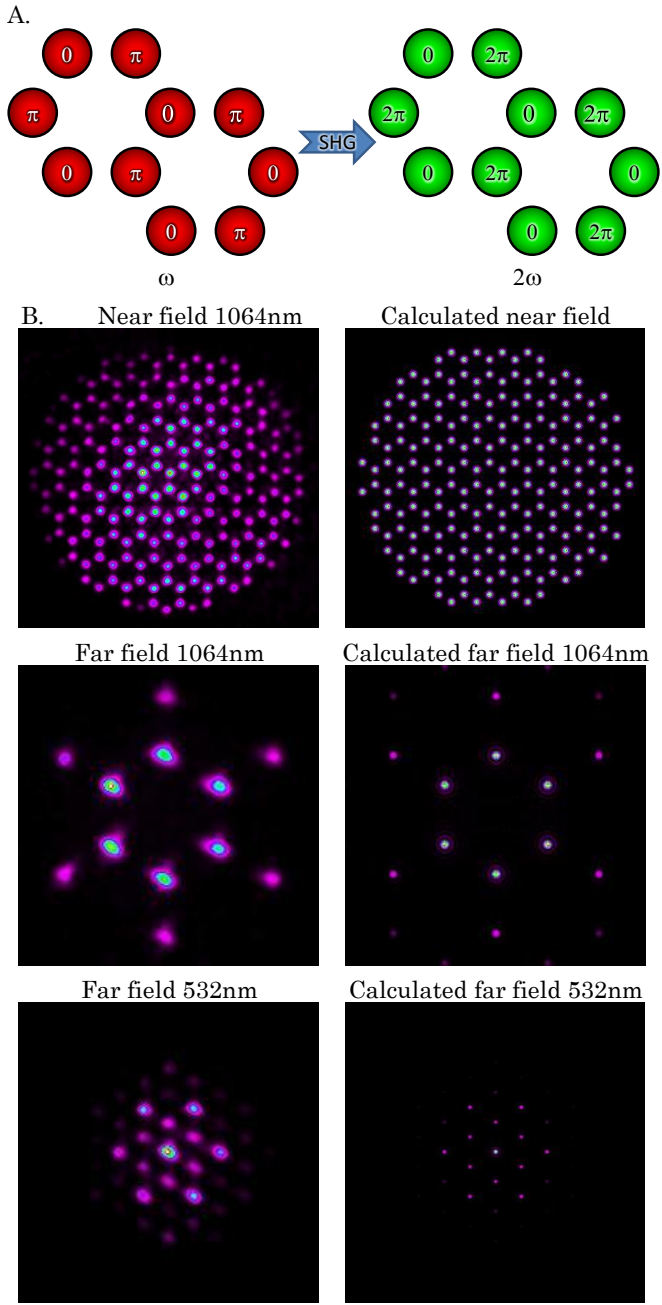

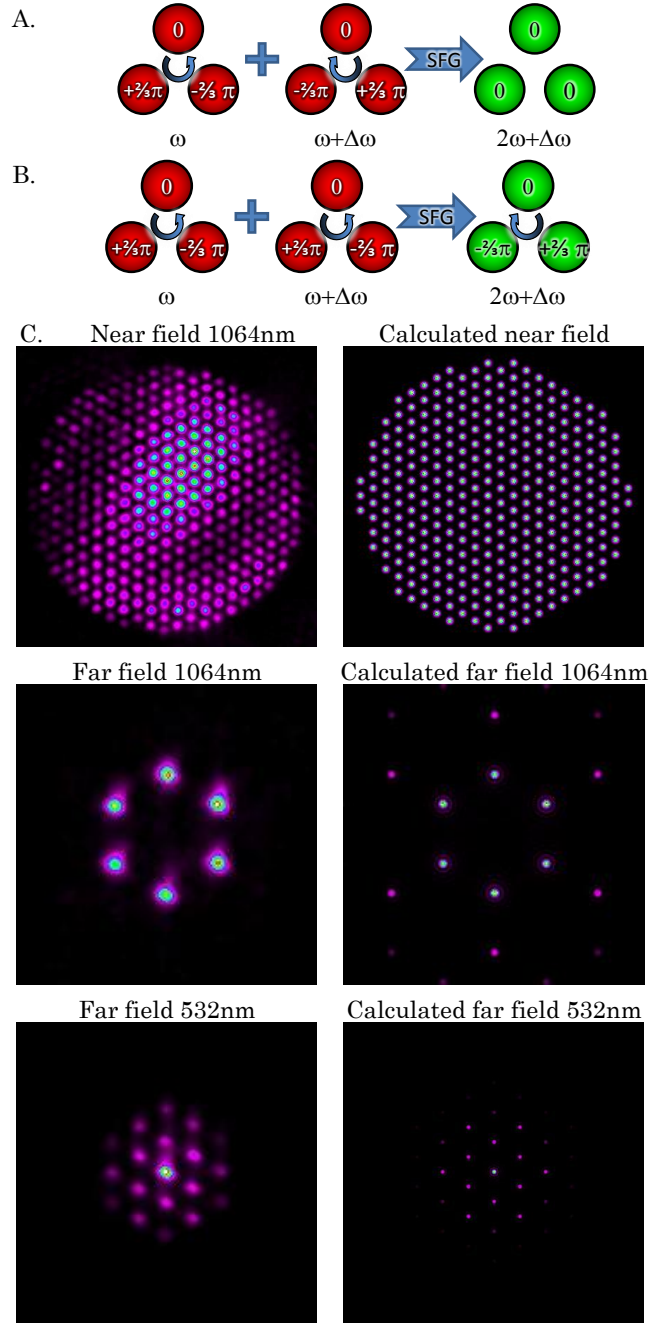

indicate again that the distance between the lobes in the far field intensity distribution in the second harmonic is half that in the first harmonic. Also, the darkness in the center in the first harmonic indicates out-of-phase locking. However, in the second harmonic, both in-phase and out-of-phase intensity distributions simultaneously appear. Finally, the sharp peaks in the far field intensity distributions indicate long range order without tight

Fig. 4. An array of 199 coupled lasers formed in a honeycomb geometry. A. Schematic illustration of phase shift doubling in the second harmonics, where an out-of-phase locking ($\pi$ phase shift) in the first harmonic is converted into an in-phase locking ($2\pi$ phase shift) in the second harmonic. B. Experimental (left) and calculated (right) intensity distributions for the near and far fields of first and second harmonics.

image of the second harmonic will simply be an inverted image of the first harmonic.

The illustrations for summing two longitudinal modes with either the same or opposite helicity, and the experimental and calculated intensity distributions for the near and far fields of first and second harmonics are presented in Fig. 5. Summing two longitudinal modes with opposite helicity yields an in-phase distribution, whereas summing two longitudinal modes with the same helicity yields an out-of-phase distribution. The intensity distributions

Fig. 5. An array of 299 coupled lasers formed in triangular geometry, each laser with many longitudinal modes. A. Schematic illustration where summing of two longitudinal modes with vortex and antivortex phase distributions leads to an in-phase intensity distribution. B. Summing two longitudinal modes both with vortex phase distributions leads to antivortex phase distribution. C. Experimental (left) and calculated (right) intensity distributions for the near and far fields of first and second harmonics.

focusing in the first harmonic, but with tight focusing in the second harmonic.

The reason for the combined in-phase and out-of-phase intensity distributions in the second harmonic is that our laser cavity contains many longitudinal modes (about 100), and each randomly has either a vortex or antivortex phase distribution. When performing a second order nonlinear process such as SHG, many new longitudinal modes are generated by summation of any two longitudinal modes in the first harmonic, namely sum frequency generation (SFG) [28]. Statistically, 50% of the energy should be in the in-phase intensity distribution, which is in very good agreement to the experimentally measured value of 47%. Our SFG results thus reveal that many longitudinal modes with uncorrelated phase distribution exist in a triangular array of lasers. Such a revelation is not possible with first harmonics only.

## 4. CONCLUSION

To conclude, we have shown, both experimentally and numerically, that second harmonics can be exploited to convert an array of out-of-phase locked lasers into in-phase locked lasers so as to allow for tight focusing, and to investigate unusual effects of certain array geometries. Our experimental arrangement is relatively simple and can easily be adapted to various arrays of negatively coupled lasers. We also found that as a result of sum frequency generation, the energy from an array of lasers arranged in a triangular geometry with vortex and antivortex phases can reach 50% in the in-phase intensity distribution. In our experiments, the efficiency of the SHG was low, but we expect that it can be significantly increased by resorting to pulse laser operations [28] and intra-cavity non-linear crystals [34–36]. In the future, we plan to use second and third harmonics as tools to investigate arrays of negatively coupled lasers formed in kagome geometry and one dimensional closed ring geometry.

## ACKNOWLEDGMENTS

The work was supported in part by the Minerva Foundation and ISF Bikura foundation. The author would like to thank Avi Pe'er for helpful suggestions on high harmonics in laser arrays.